\documentclass[floats,floatfix,amssymb,prd,twocolumn,superscriptaddress,nofootinbib,reprint]{revtex4-2}

\usepackage{natbib}
\usepackage{amssymb,amsmath,verbatim,mathtools,needspace,enumitem,etoolbox,graphicx,physics,microtype,afterpage,xspace,tabularx,lmodern,multirow,bm}
\usepackage{gensymb}
\usepackage{microtype} 
\usepackage[dvipsnames, usenames]{xcolor}
\definecolor{linkcolor}{rgb}{0.0,0.3,0.5}
\usepackage[unicode, colorlinks=true, linkcolor=linkcolor, citecolor=linkcolor, filecolor=linkcolor, urlcolor=linkcolor, linktocpage, breaklinks]{hyperref}
\usepackage[all]{hypcap}
\usepackage[T1]{fontenc}
\usepackage[utf8]{inputenc}
\usepackage[usenames,dvipsnames]{xcolor}
\usepackage{csquotes}
\usepackage{booktabs}
\hypersetup{colorlinks=true,citecolor=romared,linkcolor=romared,urlcolor=romared}

\definecolor{romared}{RGB}{142,0,28}
\usepackage{appendix}

\usepackage[normalem]{ulem}

\begin{document}
\title{Chaotic signatures in extreme-mass-ratio systems surrounded by bosonic environments}
\author{Qi-Xuan Xu}
\affiliation{CENTRA, Departamento de Física, Instituto Superior Técnico – IST, Universidade de Lisboa – UL, Avenida Rovisco Pais 1, 1049-001 Lisboa, Portugal}
\author{Kyriakos Destounis}
\affiliation{CENTRA, Departamento de Física, Instituto Superior Técnico – IST, Universidade
de Lisboa – UL, Avenida Rovisco Pais 1, 1049-001 Lisboa, Portugal}
\affiliation{School of Applied Mathematical and Physical Sciences – SEMFE, Physics Department, National Technical University of Athens -- Polytechnic Campus, Zografou 15772, Athens, Greece}
\author{Yin-Da Guo}
\affiliation{Key Laboratory of Particle Physics and Particle Irradiation (Ministry of Education),\\Institute of Frontier and Interdisciplinary Science, Shandong University, Qingdao 266237, China}
\author{Richard Brito}
\affiliation{CENTRA, Departamento de Física, Instituto Superior Técnico – IST, Universidade de Lisboa – UL, Avenida Rovisco Pais 1, 1049-001 Lisboa, Portugal}
\author{Taillte May}
\affiliation{CENTRA, Departamento de Física, Instituto Superior Técnico – IST, Universidade de Lisboa – UL, Avenida Rovisco Pais 1, 1049-001 Lisboa, Portugal}
\begin{abstract}
Extreme-mass-ratio inspirals are key gravitational-wave sources for testing strong gravity and the environments of massive black holes. While orbital motion in Kerr black hole spacetimes is integrable, realistic black holes may be surrounded by environments that break integrability. We investigate conservative orbital dynamics around rotating black holes surrounded by scalar clouds using fully nonlinear numerical solutions and an analytical post-Newtonian model. Poincaré sections and rotation curves reveal finite-width resonant islands, providing evidence for non-integrability even when the cloud carries only a small fraction of the black-hole mass. We show that the post-Newtonian description reproduces the dominant resonance and its island width for weak clouds, but underestimates the width in the configurations with stronger, more compact clouds examined here. Moreover, we find that for a scalar cloud as massive as the black hole, the phase space exhibits wider resonant islands as well as scattered layers indicative of chaotic motion. Together with earlier work, these results show that environmental effects can generically break the integrability of geodesic motion, motivating further study of resonance crossings and their gravitational-wave signatures once radiation reaction is included.
\end{abstract}
\maketitle
%
\section{Introduction}
Gravitational-wave (GW) observations provide access to the dynamics of compact objects in strong gravitational fields. Future, space-based interferometers, such as the Laser Interferometer Space Antenna (LISA)~\cite{LISA:2024hlh}, TianQin~\cite{Luo:2025ewp}, and Taiji~\cite{Luo:2021qji} will extend the observational window to frequencies that are much lower than the current kHz ground-based observations. One of the most exciting targets of these detectors are extreme-mass-ratio inspirals (EMRIs). An EMRI consists of a stellar-mass compact object slowly inspiraling into a supermassive black hole (BH)~\cite{Amaro-Seoane:2007osp,Amaro-Seoane:2012lgq,Cardenas-Avendano:2024mqp}, while producing mHz GWs. Because the signal remains in band for many orbital cycles, even small changes in the dynamics that accumulate in the GW phase are potentially detectable. This makes EMRIs sensitive probes of the central spacetime geometry~\cite{Ryan:1995wh,Moore:2017lxy,Destounis:2023khj} and of matter surrounding the primary~\cite{Barausse:2014tra,Cardoso:2021wlq,Polcar:2022bwv,Cardoso:2022whc,Destounis:2022obl,Dyson:2025dlj,Destounis:2025tjn,Copparoni:2025vty,Duque:2025yfm,HegadeKR:2025rpr,Dyson:2026ddd, Lui:2026uai}. Understanding how environmental effects shape EMRI dynamics is therefore essential, both for extracting astrophysical information and for distinguishing these effects from genuine departures from general relativity.

Ultralight bosonic fields provide a theoretically motivated class of BH environments~\cite{Brito:2025ojt}. Around BHs, they admit quasi-bound state solutions characterized by a complex oscillation frequency $\omega=\omega_R+i\omega_I$~\cite{Detweiler:1980uk,Dolan:2007mj}. A quasi-bound state with azimuthal number $m$ can extract rotational energy and angular momentum from a spinning BH when the superradiant condition
$\omega_R<m\Omega_H$
is satisfied, where $\Omega_H$ is the angular velocity of the BH horizon~\cite{Zeldovich:1971ffh,Press:1972zz,Detweiler:1980uk,Dolan:2007mj,Brito:2015oca}. Because these states are confined, their amplitudes grow exponentially, forming a macroscopic scalar cloud of mass $M_c$~\cite{Arvanitaki:2010sy,Brito:2014wla,East:2017ovw}. In the absence of competing processes, the extraction of angular momentum spins down the BH until the superradiant threshold $\omega_R=m\Omega_H$ is reached and the instability saturates~\cite{Brito:2014wla,East:2017ovw}.
The efficiency of this process depends strongly on how the boson's reduced Compton wavelength, $\hbar/(m_b c)$, compares with the BH gravitational radius, $GM/c^2$, where $m_b$ is the boson rest mass and $M$ is the BH mass. The ratio of the latter to the former defines the dimensionless gravitational coupling, $\alpha \equiv GMm_b/(\hbar c)$. Superradiant growth is exponentially suppressed for $\alpha \gg 1$~\cite{Zouros:1979iw}, whereas for $\alpha \ll 1$ it is suppressed by a high power of $\alpha$~\cite{Detweiler:1980uk, Bao:2022hew}. The process is therefore most efficient for $\alpha \sim \mathcal{O}(0.1\text{--}1)$, when the boson's Compton wavelength is comparable to the size of the BH~\cite{Dolan:2007mj}. Therefore, the more massive the BH the lighter the boson for which the process could be relevant.

The impact of boson clouds on EMRIs has only recently begun to be studied, using either Newtonian approximations (see, e.g.,~\cite{Ferreira:2017pth,Hannuksela:2018izj,Baumann:2019ztm,Tomaselli:2023ysb,Tomaselli:2024bdd,Boskovic:2025ixx}) or relativistic perturbation theory~\cite{Duque:2023seg,Brito:2023pyl,Dyson:2025dlj,Li:2025ffh,Keijzer:2026vul,Xu:2026cky,Xu:2026aic}. In particular, relativistic studies of nonspherical clouds around spinning BHs have focused mainly on dissipative effects, while the impact of the cloud's gravitational field on generic, eccentric and inclined orbits remains largely unexplored.
A natural starting point for closing this gap is provided by exact solutions describing such systems. At the superradiant threshold, Einstein's equations minimally coupled to complex massive bosonic fields admit stationary, axisymmetric BH solutions that describe bosonic clouds co-rotating with the central BH~\cite{Herdeiro:2014goa,Herdeiro:2016tmi,May:2024npn}. Geodesic motion of massive test particles in these spacetimes has been studied before, but only for circular equatorial orbits~\cite{Collodel:2021jwi,Delgado:2023wnj}.

In the Kerr spacetime, geodesic motion of massive test particles admits four constants of motion: the particle's energy, its azimuthal angular momentum, the Carter constant, and the normalization of its four-velocity~\cite{Carter:1968rr}. Since there are as many constants as degrees of freedom, the motion is Liouville-integrable and does not exhibit chaotic behavior. On the other hand, for generic stationary and axisymmetric spacetimes, the existence of a Carter-like constant that ensures integrability is not guaranteed. When no such constant exists, the radial and polar motion must be studied without assuming separability of the geodesic equations. The hidden symmetry associated with the existence of the Carter constant can, for example, be broken when the multipolar structure of Kerr is perturbed~\cite{Destounis:2020kss,Destounis:2021mqv,Destounis:2021rko,Destounis:2023cim,Destounis:2023gpw,Destounis:2023khj,Eleni:2024fgs,Destounis:2025tjn}, or when the central compact object is tidally deformed by a tertiary~\cite{Destounis:2026tet}. However, departures from Kerr do not necessarily destroy integrability, and certain non-vacuum geometries can retain separable geodesic equations~\cite{Kocherlakota:2025cwq}. 

Importantly, non-integrability can lead to chaotic motion~\cite{Contopoulos_book}. Nevertheless, its observational relevance does not require strongly chaotic trajectories: signatures of non-integrability can also arise near orbital resonances, where resonant orbits form islands of stability~\cite{Apostolatos:2009vu,Lukes-Gerakopoulos:2010ipp}. Transient orbital resonances have been shown to affect EMRI evolution even in vacuum Kerr spacetimes, where geodesics are integrable, and can introduce significant dephasing~\cite{Flanagan:2010cd,Berry:2016bit,Speri:2021psr,Lynch:2024ohd,Levati:2025ybi}. In non-integrable systems, orbits inside resonant islands are phase-locked, meaning that the fundamental frequencies describing the motion remain in a fixed rational ratio \cite{Apostolatos:2009vu,Lukes-Gerakopoulos:2010ipp,Destounis:2026tet}. During an inspiral, passage through these islands can prolong the impact of resonances and produce characteristic features in the orbital and GW frequency evolution~\cite{Destounis:2021mqv,Destounis:2021rko}. These effects motivate examining whether boson clouds generate similar resonant structures, even when their contribution to the background geometry is small. 

Motivated by this, we investigate 
geodesic dynamics in rotating BH--boson cloud spacetimes using both an analytical post-Newtonian (PN) approximation of the geometry as well as fully-nonlinear (FN), i.e., non-perturbative, numerical solutions~\cite{Herdeiro:2014goa,May:2024npn}. In both cases, we find resonant islands of finite width, providing clear evidence of non-integrability even when the cloud carries a relatively small fraction of the spacetime's total mass. Moreover, clouds with masses comparable to that of the BH exhibit multiple wide resonant islands, as well as signatures of fully chaotic motion. These results imply that waveform models for EMRIs on generic orbits in such environments will need to carefully account for orbital resonance crossings.

The paper is organized as follows. In Sec.~\ref{Sec:Geodesics and chaos}, we introduce the geodesic equations and the diagnostics of non-integrability. In Sec.~\ref{Sec:Kerr black holes with synchronized scalar hair}, we describe the synchronized BH--cloud solutions and analyze orbital dynamics in these geometries. In Sec.~\ref{Sec:Post-Newtonian metric}, we derive an analytical PN metric for the BH--cloud system and compare the resulting resonant structure of geodesics with the results in the FN solutions. Finally, in Sec.~\ref{Sec:Discussion and Conclusions}, we discuss the implications and limitations of our results. A list of the FN configurations considered here, as well as convergence studies can be found in App.~\ref{app:geodesic_convergence}, while additional details on the PN solution are given in App.~\ref{app:pn_potentials} and~\ref{app:cloud_mass_matching}. Throughout this work, we use geometrized units, $G=c=1$, and adopt a mostly-plus metric signature.

\section{Geodesics and chaos}\label{Sec:Geodesics and chaos}

\subsection{Conserved quantities and reduced dynamics}

A stationary and axisymmetric spacetime metric admits coordinates in which the line element takes the form
\begin{equation}\label{line_element}	ds^2=g_{tt}dt^2+2g_{t\varphi}dtd\varphi+g_{rr}dr^2+g_{\theta\theta}d\theta^2+g_{\varphi\varphi} d\varphi^2,
\end{equation}
where all metric components depend only on $r$ and $\theta$. Neglecting radiation reaction, a test particle of mass $m_p$ follows a timelike geodesic,
\begin{equation}\label{geodesic_equation}
\ddot{x}^\kappa+\Gamma^\kappa_{\lambda\nu}
\dot{x}^\lambda\dot{x}^\nu=0,
\end{equation}
where a dot denotes differentiation with respect to proper time $\tau$. Stationarity and axisymmetry imply conservation of the energy $E$ and azimuthal angular momentum $L_z$:
\begin{equation}\label{energy_momentum}
\begin{aligned}
-E/m_p=g_{tt}\dot t+g_{t\varphi}\dot\varphi,\quad
L_z/m_p=g_{t\varphi}\dot t+g_{\varphi\varphi}\dot\varphi.
\end{aligned}
\end{equation}
Solving Eqs.~\eqref{energy_momentum} with respect to $\dot t$ and $\dot\varphi$ leads to two first-order decoupled ordinary differential equations
\begin{equation}\label{eq:t_phi_dot}
\begin{aligned}
\dot t=\frac{E g_{\varphi\varphi}+L_zg_{t\varphi}}
{m_p(g_{t\varphi}^2-g_{tt}g_{\varphi\varphi})},\quad
\dot\varphi=-\frac{E g_{t\varphi}+L_zg_{tt}}
{m_p(g_{t\varphi}^2-g_{tt}g_{\varphi\varphi})}.
\end{aligned}
\end{equation}
The normalization of the four-velocity, $g_{\mu\nu}\dot x^\mu\dot x^\nu=-1$, provides the constraint
\begin{equation}\label{constraint_equation}
\dot r^2+\frac{g_{\theta\theta}}{g_{rr}}\dot\theta^2
+V_{\rm eff}=0,
\end{equation}
with
\begin{equation}
V_{\rm eff}=\frac{1}{g_{rr}}\left[
1+\frac{g_{\varphi\varphi}E^2+2g_{t\varphi}EL_z+g_{tt}L_z^2}
{m_p^2(g_{tt}g_{\varphi\varphi}-g_{t\varphi}^2)}\right].
\end{equation}
Outside any event horizon, $g_{rr}$ and $g_{\theta\theta}$ are positive, so Eq.~\eqref{constraint_equation} can only be satisfied where $V_{\rm eff}\leq 0$. This condition defines the region of the $(r,\theta)$ plane accessible to a massive particle, whose boundary, $V_{\rm eff}=0$, is the curve of zero velocity (CZV), where $\dot r=\dot\theta=0$. The accessible region does not, however, distinguish bound orbits from plunging or escaping ones.

For fixed $E$ and $L_z$, the remaining dynamics involves the motion along the radial and polar directions. In a Kerr spacetime, the Carter constant provides an additional conserved quantity associated with a rank-two Killing tensor, allowing the geodesic equations to be separated \cite{Carter:1968rr}. For the BH--cloud spacetimes considered below, we evolve the radial and polar equations without imposing an additional conserved quantity, and use the resulting phase-space structure to investigate integrability.

\subsection{Numerical spacetimes and orbital evolution}

The numerical BH--cloud spacetimes we consider here are written in Lewis--Papapetrou-type coordinates~\cite{May:2024npn}:
\begin{equation}\label{WP_metric}
\begin{aligned}
ds^2={}&-fNdt^2+\frac{l}{f}\bigg[
g\left(\frac{dr^2}{N}+r^2d\theta^2\right)\\
&+r^2\sin^2\theta\left(d\varphi-\frac{\omega_{\rm m}}{r}dt\right)^2
\bigg],
\end{aligned}
\end{equation}
where $N=1-r_H/r$ and the event horizon is at $r=r_H$. The functions $(f,g,l,\omega_{\rm m})$ are functions of $r$ and $\theta$ only. The construction of these solutions is detailed in Ref.~\cite{May:2024npn} and briefly reviewed below (see also Refs.~\cite{Destounis:2023khj,Destounis:2025tjn} for related studies of geodesic motion in numerically constructed spacetimes).

We parameterize the numerical trajectories by the coordinate time $t$. For the future-directed exterior orbits considered here, $\dot t>0$, and the radial and polar velocities are
\begin{equation}
\frac{dr}{dt}=\frac{\dot r}{\dot t},\qquad
\frac{d\theta}{dt}=\frac{\dot\theta}{\dot t},
\end{equation}
with $\dot t$ determined by Eq.~\eqref{eq:t_phi_dot}. This reparameterization preserves the trajectories and their sequence of crossings through a surface of section. 
The test particle mass $m_p$ enters only through the quantities $E/m_p$ and $L_z/m_p$, and we therefore set $m_p=1$ in the numerical integrations. For the values of $E/m_p$ and $L_z/m_p$ that we consider, the geodesics span eccentricities that range from $\sim0.3$ to $\sim 0.8$, which covers a significant domain of astrophysically-relevant orbits. For fixed $E$ and $L_z$, each orbit starts at $\theta(0)=\pi/2$ with $\left.dr/dt\right|_{t=0}=0$. Varying $r(0)$ generates a family of initial conditions, with the positive initial value of $d\theta/dt$ determined by the coordinate-time form of the constraint Eq.~\eqref{constraint_equation}. This constraint also provides a consistency check during the evolution. Across the simulations, the constraint is maintained to an accuracy ranging from approximately one part in $10^6$ to one part in $10^9$ over the first $\sim10^4$ orbital revolutions. The numerical sensitivity of the measured quantities relevant to this study, with respect to the background's resolution and integration accuracy is examined in App.~\ref{app:geodesic_convergence}.

\subsection{Non-integrability, resonant islands, and chaos}

Generic bound geodesics in an integrable BH spacetime are characterized by three fundamental frequencies: the radial libration frequency $\omega_r$, describing the oscillating motion rate from periapsis to apoapsis and back; the polar libration frequency $\omega_\theta$, describing oscillations from the maximum to the minimum polar angle and back, across the equatorial plane; and the azimuthal revolution frequency $\omega_\varphi$, describing the motion around the symmetry axis~\cite{Schmidt:2002qk}. 
Moreover, due to stationarity and axisymmetry, the motion effectively reduces to a four-dimensional phase space spanned by $(r,\theta)$ and their conjugate momenta. Within this phase-space, the normalization of the four-velocity further restricts the motion to a three-dimensional hypersurface. In the integrable case, the additional Carter-like constant additionally confines a generic bound trajectory to a two-dimensional invariant torus~\cite{Contopoulos_book,lichtenberg2013regular}.

The ratio $\omega_r/\omega_\theta$ distinguishes non-resonant from resonant motion in this phase space. If the ratio is an irrational number, the radial and polar oscillations never repeat in exactly the same manner and the trajectory densely covers the surface of its invariant torus. On the other hand, if the ratio is a rational number, the two oscillation frequencies are commensurate and the reduced motion repeats after a finite time. Such a periodic trajectory samples only part of the resonant torus and does not cover it densely~\cite{Lukes-Gerakopoulos:2021ybx}. We use the term orbital resonance, or just resonance, to refer, from now on, to the radial--over--polar frequency proportionality.

The properties of generic and resonant geodesics can be visualized with a Poincar\'e surface of section. Instead of following the entire trajectory, one records its intersections through a chosen surface, such as the equatorial plane, always selecting the same crossing direction. A non-resonant orbit produces points that trace a closed invariant curve, whereas a resonant orbit produces a finite set of periodic points on the section. In the integrable case, the corresponding resonant initial conditions have zero width along the chosen family~\cite{Lukes-Gerakopoulos:2021ybx}. Nevertheless, there is no distinction when plotting non-resonant trajectories and orbits close to the resonance. The distinction only appears when fine-tuned initial conditions are used that place the orbit exactly at the invariant resonant torus, where the orbit is perfectly periodic. Plotting the intersections of many orbits gives a Poincar\'e map. In a regular region, the curves obtained this way form nested families around a central fixed point, and the frequencies vary smoothly between neighboring tori. The map therefore provides a way to examine the organization of orbital motion without trying to find an explicit additional constant of motion.

When a weak perturbation breaks integrability, regular and irregular motion can co-exist. The Kolmogorov--Arnold--Moser (KAM) theorem states that most invariant tori survive a sufficiently small perturbation, provided that the required non-degeneracy and non-resonance conditions hold \cite{Moser:430015,Arnold_1963}. Here the non-resonance condition requires the frequency ratio to be sufficiently poorly approximated by rational numbers \cite{Destounis:2020kss}. These tori are deformed by the perturbation but continue to support generic motion; in this case the orbits and their section curves are called KAM tori and KAM curves, respectively. Thus, a non-integrable system can still contain extensive regions of smooth invariant curves, that nest around a common central fixed point of the map, with departures from integrability most apparent near resonances.

Near a resonance, a generic weak perturbation can break the continuous family of periodic orbits on the resonant torus into isolated stable and unstable periodic points on the section. The alternating arrangement of these points is associated with the Poincar\'e--Birkhoff theorem \cite{Birkhoff:1913,Contopoulos_book}. Trajectories around the stable resonant points trace nested curves that form resonant islands. These islands, together with the intervening unstable periodic points, form a Birkhoff chain. An orbit on a curve in the chain visits the islands successively, tracing a curve around each stable point. Although they also oscillate about the stable periodic orbit at the island centers, the mean radial-to-polar frequency ratio remains locked to the same rational value. This phase locking extends across a finite region of initial conditions. By comparison, an isolated resonant torus of the unperturbed integrable system has no finite width across the surrounding family of tori.

The unstable periodic points are associated with chaotic layers near the island boundaries. The trajectories approaching and departing from an unstable periodic orbit define its stable and unstable manifolds; transverse intersections of these manifolds can produce chaotic motion. Such trajectories intersect the Poincar\'e section in scattered sets of points rather than smooth closed curves, and the corresponding chaotic layers can be extremely thin for weak perturbations. The periodic unstable orbit itself remains resonant, however, its instability alone does not imply chaos. The motion inside the islands and the possible chaotic motion around them are, therefore, distinct features of the resonant region \cite{Contopoulos_book,lichtenberg2013regular}. Although KAM and Poincar\'e--Birkhoff theorems explain the standard generic near-integrable picture, for a particular spacetime, the actual orbital structure must be established from its dynamics.

Another useful tool to study integrability is the rotation number. This quantity provides a quantitative way to locate the resonant islands in a Poincar\'e map \cite{Apostolatos:2009vu,Lukes-Gerakopoulos:2010ipp,Destounis:2020kss}. For a given trajectory, let $\vartheta_i$ denote the angle swept between consecutive section points about the central fixed point of the main family of invariant curves. With a consistent choice of rotation direction and angular branch, the estimate from $\mathcal N$ angular advances is
\begin{equation}\label{rotation}
\nu_\vartheta=\frac{1}{2\pi\mathcal N}
\sum_{i=1}^{\mathcal N}\vartheta_i.
\end{equation}
The rotation number then measures the average fraction of a turn made by the section points per crossing. In the limit $\mathcal N\rightarrow\infty$, the convention adopted here gives $\nu_\vartheta=\omega_r/\omega_\theta$ for the orbital families considered. Evaluating this quantity for a sequence of initial conditions gives a rotation curve: each point on that curve represents a separate trajectory with different initial conditions.

For example, one can vary the initial radius while keeping $E$, $L_z$, the initial polar angle, and the initial radial velocity fixed, with the remaining velocity set to be defined by the normalization constraint. When this sequence crosses a resonant island, the rotation number stays at the same rational value over a finite interval of initial radii. Therefore, all geodesics inside a resonant island share the same commensurability relation $\omega_r/\omega_\theta$. In this case, the rotation curve then contains a horizontal segment, called a plateau. Its width is the difference between the initial radii at the two endpoints and measures the width of the island along the chosen initial-condition line. A resolved island chain and its corresponding plateau provide complementary evidence for non-integrability. Near unstable periodic orbits and chaotic layers, the rotation curve may instead show irregular variations. 

Resonances also matter for the slow evolution of an EMRI under radiation reaction. Even in integrable Kerr backgrounds, passage through a radial--polar resonance changes how the orbit samples its radial and polar phases. Contributions to the self-force that normally average out can accumulate, producing phase-dependent changes in the orbital evolution and GW phase \cite{Flanagan:2010cd,Flanagan:2012kg}. The treatment of orbital resonances have consequently received considerable attention \cite{Flanagan:2010cd,Flanagan:2012kg,Brink:2013nna,Ruangsri:2013hra,Brink:2015roa,Berry:2016bit,Bonga:2019ycj,Speri:2021psr,Levati:2025ybi,Bonga:2026mmz}. In a non-integrable background, a finite-width island can support an extended interval of phase locking as the secondary crosses through it. Approximate inspiral calculations have found such occurrences to last hundreds of orbital cycles, including examples with approximately $200-300$ cycles \cite{Contopoulos:2011dz,Destounis:2021mqv}, and to produce abrupt features in the GW frequency evolution \cite{Destounis:2021mqv,Destounis:2021rko}. These durations depend on the system parameters, initial conditions, and radiation-reaction prescription. Extended trapping is also distinct from a sustained resonance lasting as long as the inspiral timescale \cite{vandeMeent:2013sza}.
A wider island permits a longer resonant occurrence. However its width alone, does not determine the time the system remains on resonance. A more complete description requires following the inspiral through the evolving phase space, including its evolution before and after the resonance~\cite{Ruangsri:2013hra, Berry:2016bit, Speri:2021psr}, and accounting for the relevant self-force contributions~\cite{Barack:2009ux}.

In what follows, we use Poincar\'e maps and rotation curves to search for signatures of non-integrability in the geodesic motion around rotating BHs surrounded by scalar clouds. We record $(r/M,dr/dt)$ at each subsequent equatorial crossing with $d\theta/dt>0$ to produce Poincar\'e maps for the initial conditions discussed above. The same family of initial conditions is used to define the rotation curves. 

\section{Kerr black holes with synchronized scalar hair}\label{Sec:Kerr black holes with synchronized scalar hair}

The action for a complex scalar field $\Phi(x^\mu)$ with mass $m_b\equiv \mu \hbar$, minimally coupled to gravity is given by
\begin{equation}
\mathcal{S}\equiv \int d^4x \sqrt{-g}\left(\frac{R}{16 \pi}- g^{\kappa\nu}\partial_{\kappa}\Phi\partial_{\nu}\Phi^*-\mathcal{U}\right)\,,\label{eq:action}
\end{equation}
where $R$ is the Ricci scalar of the spacetime metric $g_{\kappa \nu}$, $g\equiv \det(g_{\kappa \nu})$ is the metric determinant, and $\mathcal{U}(|\Phi|^2)$ is a real-valued, $U(1)$-invariant, self-interaction potential. For a weak scalar field $|\Phi| \ll 1$, the self-interaction potential is $\mathcal{U}\sim \mu^2|\Phi|^2+{\cal O} (|\Phi|^4)$. Our methods are applicable, in principle, to any nonlinear potential, though here we only consider the second-order term.

Varying this action with respect to the metric and the scalar field, we get the Einstein-Klein-Gordon (EKG) field equations:
\begin{equation}
\begin{aligned}\label{eq: EKG field equation}
    {G}_{\kappa\nu} = 8\pi T^{{\Phi}}_{\kappa\nu},\qquad
    \Box {\Phi} = \mu^2 {\Phi},
\end{aligned}
\end{equation}
where ${G}_{\kappa\nu}:= {R}_{\kappa\nu}-{g}_{\kappa\nu}{R}/2$ and $\Box:={g}_{\kappa\nu}\nabla^\kappa\nabla^{\nu}$.  
The stress-energy tensor of the scalar field, $T_{\kappa\nu}^{\Phi}$, reads
\begin{equation}\label{eq: stress-energy tensor definition}
T_{\kappa\nu}^{\Phi}=2\partial_{(\kappa}{\Phi}\partial_{\nu)}{\Phi}^*-{g}_{\kappa\nu}\left(\partial_\lambda{\Phi}\partial^\lambda{\Phi}^*+\mu^2|{\Phi}|^2\right).
\end{equation}

The FN numerical backgrounds used in this work are stationary and axisymmetric solutions of the full EKG equations~\eqref{eq: EKG field equation}, constructed with the code of Ref.~\cite{May:2024npn}; solutions of this kind were first obtained in Ref.~\cite{Herdeiro:2014goa}. In contrast to a perturbative treatment of the scalar field on a fixed Kerr spacetime, by construction, the gravitational backreaction of the cloud on the geometry is fully taken into account in these solutions. 

For the resulting spacetime to be stationary and axisymmetric, we impose the following harmonic dependence ansatz for complex scalar field:
\begin{equation}
\Phi(t,r,\theta,\varphi)
=
\phi(r,\theta)e^{-i\omega_R t+i m\varphi}.
\end{equation}
Regularity at the BH horizon also requires the synchronization condition,
\begin{equation}
\omega_R=m\Omega_H\,,
\end{equation}
where $\Omega_H=-\lim_{r\to r_H}g_{t\varphi}/g_{\varphi\varphi}$. 
These solutions therefore describe BH--cloud configurations in equilibrium at the threshold of the superradiant instability, where the scalar field neither grows through superradiance nor is absorbed by the BH. In this work, we consider the fundamental $m=1$ scalar configurations of Refs.~\cite{Herdeiro:2014goa,May:2024npn}, which, in the test-field limit, connect continuously to the end state of the fastest growing dipolar scalar mode~\cite{Herdeiro:2017phl}. 

Although these solutions are themselves unstable against superradiant growth of higher-$m$ modes~\cite{Ganchev:2017uuo}, the instability timescale can be extremely long for configurations that form dynamically through superradiance, which have cloud masses of at most $M_c/M\sim0.1$~\cite{Herdeiro:2021znw}. Such configurations therefore remain astrophysically relevant~\cite{Degollado:2018ypf}. By contrast, a dynamical instability has recently been uncovered for configurations in which the cloud carries a large fraction of the spacetime's mass, causing the BH to be ejected from the cloud on a very short timescale~\cite{Nicoules:2025bhh}. It is currently unclear above which value of $M_c/M$ this instability becomes important. We nevertheless also consider a configuration with $M_c/M\sim1$, not motivated by superradiance, but to understand how geodesic motion is affected in the presence of a very massive environment.

In what follows, we study the geodesic dynamics in these BH--cloud spacetimes, considering different configurations. We first identify finite-width resonant islands using a Poincar\'e section and its corresponding rotation curve. We then examine how the width of this resonance changes across a sequence of solutions with different cloud masses. Finally, we consider a configuration with $M_c/M\simeq1$ for which the phase space contains several resonant islands and chaotic layers.

\subsection{Geodesic dynamics and evidence of non-integrability}\label{Sec:Results}

\begin{figure}
    \centering
    \includegraphics[width=\columnwidth]
    {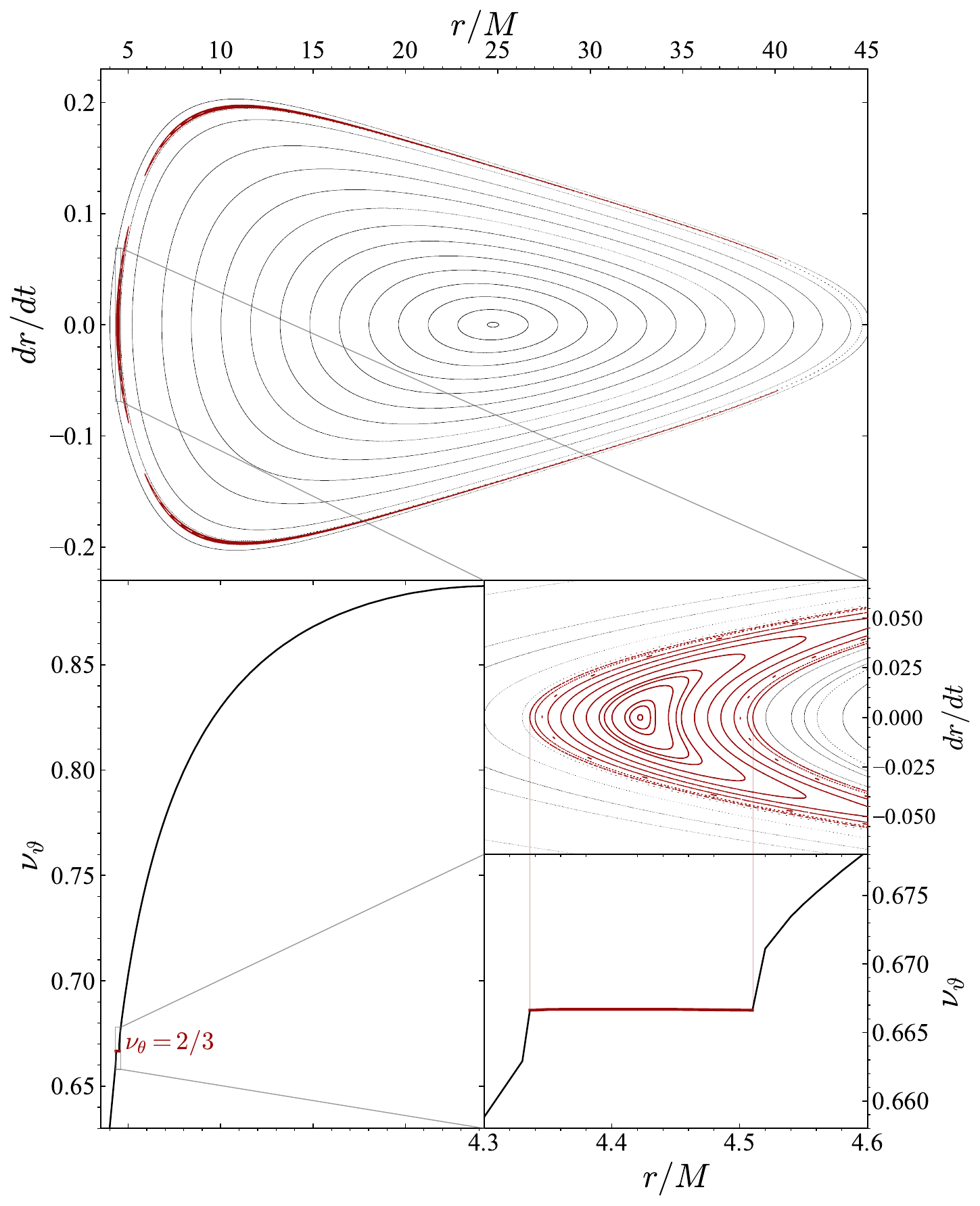}
    \caption{Equatorial Poincar\'e section and rotation curve for bound timelike geodesics with $E/m_p=0.98$ and $L_z/m_p=3.3M$ in an FN BH--cloud spacetime with $\alpha\simeq0.15$ and
    $M_c/M\simeq0.1$. \emph{Top panel:} Poincar\'e section in the $(r/M,dr/dt)$ plane. Black points trace non-resonant KAM curves, while red points identify the three-island Birkhoff chain at $\nu_\vartheta=2/3$, with nested KAM resonant curves surrounding the stable periodic points. \emph{Bottom-left panel:} Corresponding rotation curve, with the finite-width plateau at $\nu_\vartheta=2/3$ highlighted in red. \emph{Right column:} The \emph{upper} and \emph{lower}  panels magnify the resonant island and plateau shown in the top and bottom-left panels, respectively. The vertical red lines connect the plateau endpoints to the corresponding radial boundaries of the resonant island.}
    \label{fig:Nonintegrability}
\end{figure}

We begin with a FN configuration with $\alpha\simeq0.15$ and $M_c/M\simeq0.1$. Fig.~\ref{fig:Nonintegrability} combines the Poincar\'e section and the rotation curve for geodesics with $E/m_p=0.98$ and $L_z/m_p=3.3M$ obtained for this configuration. 

The top panel shows the complete Poincar\'e map. Most of the sampled orbits generate smooth closed curves around the central fixed point of the map. These curves correspond to non-resonant KAM tori that remain regular after the scalar-cloud contribution is included. Their presence is consistent with the KAM theorem.
The trajectories shown in dark red have a different structure. Their intersections are divided among three disconnected islands rather than forming one closed invariant KAM curve. These islands form a Birkhoff chain with rotation number
\begin{equation}
    \nu_\vartheta
    =\frac{\omega_r}{\omega_\theta}
    =\frac{2}{3}.
\end{equation}
The chain of three resonant islands, symmetric about the horizontal line $dr/dt=0$, arises from the break up of the $2/3$ resonant torus, as expected from the Poincar\'e--Birkhoff theorem. All geodesics inside the islands share the same resonant rotation number. 

The bottom, upper-right panel zooms into one island of this chain. The red KAM curves, which are crossed by the line $dr/dt=0$, are nested around a stable periodic point, whereas all the neighboring non-resonant KAM curves remain outside the island boundary. This is reflected in the width of the plateau in the rotation curve when $\nu_\vartheta=2/3$. We note that the island also intersects lines of nonzero $dr/dt$. However, because the Birkhoff island chain is symmetric about $dr/dt=0$, the island attains its maximum width along the zero-radial-velocity line.

The bottom-left panel of Fig.~\ref{fig:Nonintegrability} shows the full rotation-number curve as a function of the initial radius. Away from resonances, $\nu_\vartheta$ evolves monotonically with $r$. At the $2/3$ resonance, the rotation curve develops a plateau, which is clearly visible in the zoom shown in the bottom-right panel. The plateau spans the approximate interval $ 4.33\lesssim  {r}/{M}\lesssim4.51$. 
For clarity, the two bottom-right panels share the same radial axis, and vertical red lines connect the edges of the rotation-number plateau to the radial boundaries of the corresponding resonant island in the Poincar\'e section. This shows directly that the width of the plateau measures the size of the resonant island along the chosen family of initial conditions. For this configuration, the island width in coordinate radius is $r_w/M\simeq0.18$.
Our inability to identify islands associated with other resonances in the Poincar\'e section is most likely a matter of resolution. Birkhoff chains associated to $1/2$, $5/7$, $3/4$ resonances, and so on, should also be present. However, their widths are expected to become increasingly narrow for higher-order resonances and are therefore harder to resolve.

\begin{figure}
    \centering
    \includegraphics[width=\columnwidth]{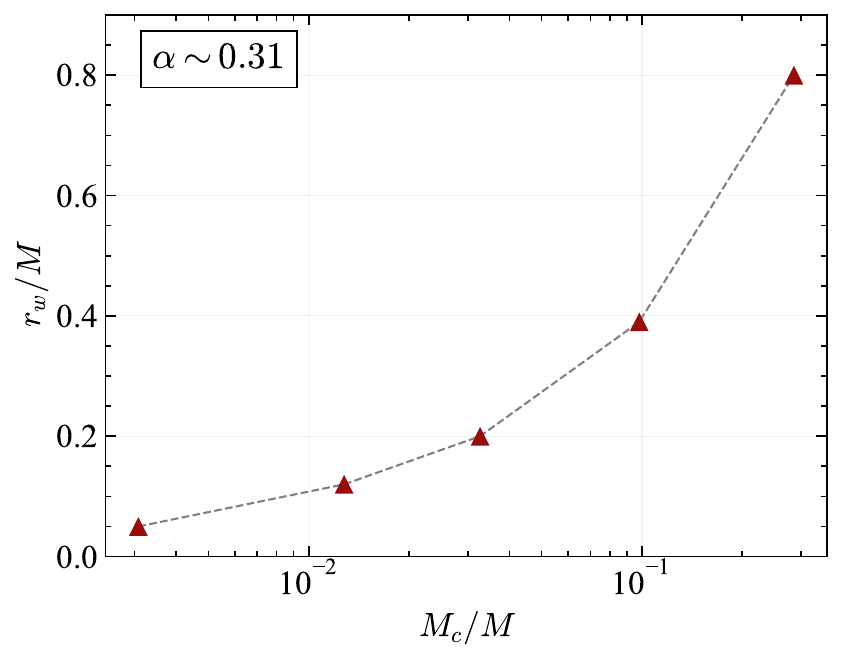}
    \caption{Width of the $2/3$ resonant island for five FN configurations spanning $0.003\lesssim M_c/M\lesssim0.3$ with $\alpha\simeq0.31$. The widths are computed for geodesics with $E/m_p=0.95$ and $L_z/m_p=2.6M$. The sampled sequence exhibits a monotonic increase in $r_w/M$ with $M_c/M$.}
    \label{fig:plateau_width}
\end{figure}

To further understand how the cloud affects the orbital motion near resonances, we compute the width of the $2/3$ resonance for five FN configurations with fixed $\alpha\simeq 0.31$ and varying cloud mass fraction $M_c/M$, which at fixed $\alpha$ controls the compactness of the cloud. This is shown in Fig.~\ref{fig:plateau_width}, where we compare the island widths normalized by the BH mass, $r_w/M$, as a function of $M_c/M$. As before, $r_w$ is determined from the separation between the two boundaries of the $2/3$ plateau along the corresponding family of initial conditions. A non-zero width is resolved for all five configurations. The lightest configuration we tested has $M_c/M\simeq 0.003$. Even though it corresponds to a cloud with a very small compactness, the measured island width is still much larger than the numerical uncertainty in locating the island boundaries. On the other hand, the heaviest configuration included in Fig.~\ref{fig:plateau_width} has $M_c/M\simeq0.3$ and gives the largest measured width. 
The five data points show a monotonic increase of $r_w/M$ over the sampled range $0.003\lesssim{M_c}/{M}\lesssim0.3$. Over this range, $M_c/M$ increases by approximately two orders of magnitude, while the measured resonance width increases by a factor of approximately $16$.

Although we could be tempted to determine a power-law dependence of $r_w$ with respect to $M_c$, we should be careful in doing so. This is due to the fact that the configurations do not differ only in the cloud mass $M_c/M$. The scalar frequency and BH angular momentum also vary across the sequence of solutions. 
Figure~\ref{fig:plateau_width}, therefore, establishes a monotonic relation for the sampled solutions, but it does not isolate the dependence on $M_c$ from the dependence on the other parameters. Nevertheless, the overall behavior of $r_w$ is consistent with earlier studies in which increasing a non-Kerr deformation enlarges the corresponding resonant island widths \cite{Apostolatos:2009vu,Lukes-Gerakopoulos:2010ipp,Destounis:2020kss,Destounis:2023cim,Destounis:2023khj,Destounis:2025tjn,Destounis:2026tet}.

In conclusion, as explained in Sec.~\ref{Sec:Geodesics and chaos}, the islands in the Poincar\'e section and the plateau in the rotation curve are two manifestations of the same effect. Their simultaneous presence in Fig.~\ref{fig:Nonintegrability} demonstrates that geodesic motion in BH--cloud spacetimes is non-integrable, which is one of the key results of this work. Similar results have also been found for BHs surrounded by dark matter halos or tidally perturbed by a third body~\cite{
Destounis:2025tjn,Destounis:2026tet}. Taken together, these results support a strong conclusion: \emph{environments around BHs can generically break the integrability of geodesic motion}.

\subsection{Comparable-mass scalar-cloud configuration}
\label{subsec:extreme_cloud}
\begin{figure}[h]
    \centering
    \includegraphics[width=\columnwidth]{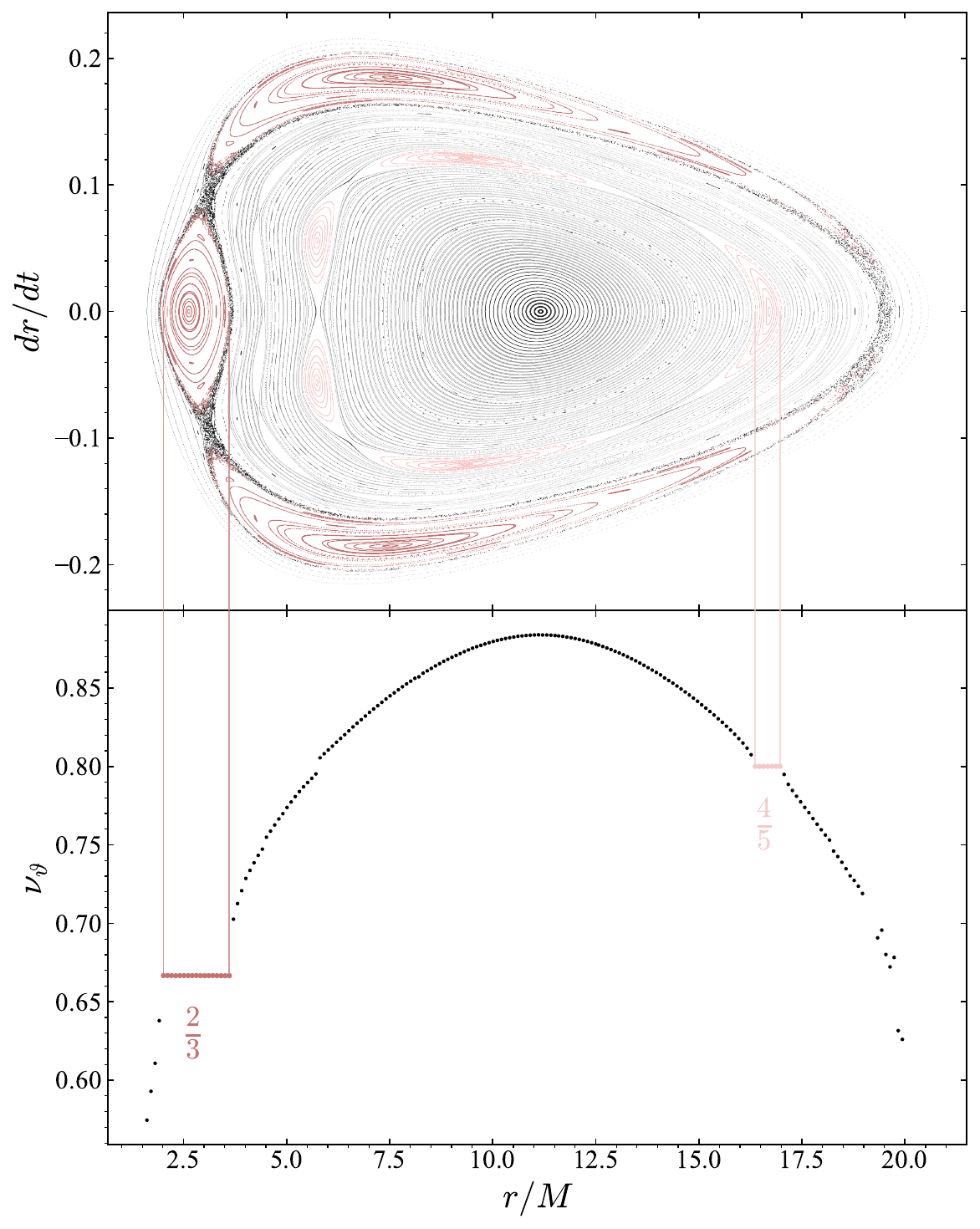}
    \caption{\emph{Top panel:} Poincar\'e section in the $\left(r/M,dr/dt\right)$ plane for the FN configuration with $\alpha\simeq0.31$, $M_c/M\simeq1$, $E/m_p=0.92$, and $L_z/m_p=2.7M$. Smooth KAM curves co-exist with a three-island chain at $\nu_\vartheta=2/3$, a five-island chain at $\nu_\vartheta=4/5$, a seven-island chain at $\nu_\vartheta=6/7$, and an eight-island chain at $\nu_\vartheta=6/8$. Chaotic layers near the unstable periodic points of the $2/3$-resonant island are also visible as scattered black dots. \emph{Bottom panel:} Corresponding rotation curve evaluated along the $dr/dt=0$ line. The colored plateaus identify the $2/3$ and $4/5$ resonances and are connected to the corresponding island chains. The discontinuous jumps that appear are due to the crossing of regions in the vicinity of an unstable periodic point, while the scattered points around $r\simeq 19.6M$ designate a clear passage through a chaotic region.}
    \label{fig:extreme}
\end{figure}

The results shown so far consider only configurations in which the cloud's contribution to the spacetime's total mass is subdominant to that of the BH. We now examine a case where the bosonic environment is as massive as the BH, by considering a FN configuration with $ M_c\simeq M$ and $\alpha\simeq0.31$.
The corresponding Poincar\'e section and rotation curve are shown in Fig.~\ref{fig:extreme}. In this case, we used $E/m_p=0.92$ and $L_z/m_p=2.7M$ in order to obtain a closed CZV that contains several low-order resonances. The central fixed point is approximately at $r\simeq 12M$.

The top panel of Fig.~\ref{fig:extreme} demonstrates that, even in this case, a large part of the densely-sampled phase space is still occupied by KAM curves. Several resonant islands can also be clearly seen. In particular, the widest island symmetric about the line $dr/dt=0$, shown in dark red, spans $2.1\lesssim r/M\lesssim3.7$ and belongs to the Birkhoff chain with rotation number $\nu_\vartheta=2/3$. We also show, in light red, a second chain of resonant islands. This resonance has multiplicity $5$, which means that the quintuplet of resonant islands corresponds to the stable periodic orbit with $\nu_\vartheta=4/5$.  
This chain is also symmetric about the line $dr/dt=0$, but here the line of initial conditions used to construct the rotation curve crosses an island on the right-hand side of the central fixed point of the Poincar\'e section, at $16.1\lesssim r/M\lesssim16.7$. Therefore, the $4/5$ plateau in the corresponding rotation curve of Fig. \ref{fig:extreme} (bottom panel) appears on the right of the largest rotation number. Nevertheless, an imprint of the $4/5$ resonance is also visible to the left of the maximum of the rotation curve, as an abrupt jump near $\nu_\vartheta=4/5$ at $r\sim6M$. This jump is an inflection point and it corresponds to a geodesic passing between two resonant islands, where an unstable periodic point creates a region of mild or strong chaos. Thus, inflections or randomly scattered rotation numbers indicate the crossing of chaotic layers. This is more evident in the range of $19.5\lesssim r/M\lesssim19.8$, where the rotation curve shows scattered values. These belong to the chaotic layer of the $2/3$ resonance, visible in the Poincar\'e section. Within such layers, the motion is purely chaotic and the rotation number is ill-defined.
 
Besides the $2/3$ and the $4/5$ resonant islands and their corresponding chaotic layers, we can also see two more resonances that leave a smaller imprint on the Poincar\'e map and rotation curve. The first chain, that corresponds to the resonance $6/8$, is crossed by the line $dr/dt=0$ near its unstable periodic points, which translates to two small inflection points on the rotation curve, visible on the left and the right hand side of the maximum rotation number. The second one that is only slightly visible is the $6/7$ chain. In this case, the line $dr/dt=0$ crosses through one of its islands to the left hand side of the central fixed point, around $r\simeq 8M$, whereas on the right it passes near an unstable periodic point. The corresponding inflection point is, however, too small to be visible in the rotation curve. 

These results suggest that the phase space becomes increasingly complex as the cloud mass increases or, equivalently, as the spacetime departs further from the Kerr geometry. 
We note, however, that the orbital parameters used in Fig.~\ref{fig:extreme} differ from those used for the lighter clouds (cf. Figs.~\ref{fig:Nonintegrability} and~\ref{fig:plateau_width}), so caution should be exercised when comparing results. 
A complete scan over $E$ and $L_z$ is computationally costly and left for future work. Nor should these results be interpreted as establishing a general threshold for the onset of strong chaotic motion. Rather, they illustrate how complex the dynamics can become when the cloud mass is comparable to that of the BH, motivating more systematic studies.

\section{Post-Newtonian metric}\label{Sec:Post-Newtonian metric}

A difficulty in studying geodesic motion in the FN BH--cloud solutions is that they are obtained numerically, which makes it hard to explore the parameter space systematically. In this section, we therefore derive an analytical approximation to the metric using a PN expansion and test it by reproducing some of the results of the previous section.

In the non-relativistic regime, $\alpha\ll1$, it is convenient to factor out the fast oscillations on the timescale $\mu^{-1}$ by adopting the ansatz:
\begin{equation}
    {\Phi}(t,r,\theta^A) = e^{-i\mu t}\hat{\Phi}(t,r,\theta^A).
\end{equation}
One can then show that, in the Newtonian limit, the EKG system~\eqref{eq: EKG field equation} reduces to the Schr\"odinger--Poisson system (see, e.g.,~\cite{Annulli:2020lyc}):
\begin{align}
    \nabla^{2}U &= 4\pi \left[M\delta(\textbf{r})+2 \mu^2|\hat{\Phi}|^{2}\right],\label{eq:Poisson equation}\\
    i\frac{\partial\hat{\Phi}}{\partial t} &= -\frac{1}{2\mu}\nabla^{2}\hat{\Phi}+\mu U\hat{\Phi},\label{eq:Shro equation}
\end{align}
where $U$ is the Newtonian potential, with $|U|\ll1$, and the BH is modeled as a point mass through the delta function. 

For a sufficiently dilute cloud, the gravitational potential is dominated at leading order by that of the BH, $U\simeq U^{\rm BH}=-M/r$. In this case, Eq.~\eqref{eq:Shro equation} reduces to the Schr\"odinger equation with a Coulomb-like potential, whose bound-state solutions take the hydrogenic form,
\begin{equation}
    \hat{\Phi}(t,r,\theta^A)=e^{-iE_{n\ell}t}R^{h}_{n\ell}(r)Y_{\ell m}(\theta^{A})\,,
    \label{eq:hydrogen-like-eigenfunction}
\end{equation}
where $Y_{\ell m}(\theta^{A})$ are spherical harmonics and the radial function is
\begin{equation}
\begin{aligned}\label{eq:hydrogenic radial function}
    R^{h}_{n\ell}(r)=&\frac{\sqrt{M_c}}{\sqrt{2}\mu}\sqrt{\left(\frac{2\mu\alpha}{\bar{n}}\right)^3\frac{(\bar{n}-\ell-1)!}{2\bar{n}(\bar{n}+\ell)!}}\left(\frac{2\alpha\mu r}{\bar{n}}\right)^{\ell}\\
    \times&L_{\bar{n}-\ell-1}^{2\ell+1}\left(\frac{2\mu\alpha r}{\bar{n}}\right)\exp\left(-\frac{\mu\alpha r}{\bar{n}}\right),
\end{aligned}
\end{equation}
with principal quantum number $\bar n=n+\ell+1$, and $L_{\bar n-\ell-1}^{2\ell+1}$ are the associated Laguerre polynomials. The energy eigenvalues in Eq.~\eqref{eq:hydrogen-like-eigenfunction} are defined by~\cite{Detweiler:1980uk}
\begin{equation}\label{eq:energy level}
    E_{n\ell} = \omega_{n \ell}-\mu\simeq -\mu\frac{\alpha^2}{2\bar{n}^2},
\end{equation}
where $\omega_{n \ell}$ are the corresponding eigenfrequencies, which in the Newtonian limit are purely real numbers.   

Notice that Eqs.~\eqref{eq:Poisson equation} and \eqref{eq:hydrogenic radial function} implicitly suggest that we have normalized the scalar field amplitude such that the cloud's total mass $M_c$ is
\begin{equation}
 2\mu^2\int d^3x\,|\hat{\Phi}|^2=M_c.
\end{equation}
The parameter $M_c$ defined in this way equals the total mass of the cloud only in the \emph{Newtonian limit}. To keep the expressions compact, we retain this notation in the PN construction and in App.~\ref{app:pn_potentials}. The relation between this parameter and the gravitational mass of the cloud in the FN solutions is discussed in App.~\ref{app:cloud_mass_matching}.

With these hydrogenic solutions, we can now derive the leading PN corrections to the BH metric due to the presence of the cloud, following a construction similar to that of Ref.~\cite{Mitra:2025tag}. The field equations obtained by varying the action~\eqref{eq:action} with respect to $\Phi^*$ and $g_{\kappa\nu}$ can be
written as
\begin{align}
    &\frac{1}{\sqrt{-g}}\partial_{\kappa}\left(\sqrt{-g}g^{\kappa\nu}\partial_{\nu}\Phi\right)= \mu^2\Phi\,, \\
&R_{\kappa\nu}= 8\pi \left(\widetilde{T}^S_{\kappa \nu }+\widetilde{T}^{\rm BH}_{\kappa \nu }\right)\,,\label{EOM_BosonSapp}
\end{align}
where the Einstein equations are written in an alternative form using the trace-reversed stress-energy tensor of the scalar field
\begin{equation}
    \widetilde{T}^S_{\kappa \nu }\equiv T^S_{\kappa \nu}-\frac{1}{2}T^S g_{\kappa \nu}=2\partial_{(\kappa}\Phi^* \partial_{\nu)}\Phi+g_{\kappa \nu}\mu^2|\Phi|^2,  
\end{equation}
and we again model the BH as a point-particle located at the origin
\begin{equation}
    \widetilde{T}^{\rm BH}_{\kappa \nu }\equiv T^{\rm BH}_{\kappa \nu}-\frac{1}{2}T^{\rm BH} g_{\kappa \nu}=M\delta(r)\delta^{0}_{\kappa}\delta^{0}_{\nu}+\frac{1}{2}M\delta(r)g_{\kappa\nu}.
\end{equation}
For the purpose of analytically deriving the cloud's corrections to metric, the
central BH is represented only by its Newtonian mass monopole,
\begin{equation}
 \nabla^2U^{\rm BH}=4\pi M\delta^{(3)}(\boldsymbol{x}),
 \qquad U^{\rm BH}=-\frac{M}{r}.
 \label{eq:BH_Newtonian_potential}
\end{equation}
This approximation is used only in the
BH--cloud interaction term discussed below. We do not use it to derive higher-PN corrections to the vacuum BH background, which we instead take it to be Kerr~\cite{Mitra:2025tag}.

In harmonic coordinates, the 1PN metric takes the form~\cite{Poisson:2014}
\begin{align}\label{eq:1pNmetric}
g_{tt}&=-1-2U -2\left(\psi+ U^2\right)+\mathcal{O}(v^6)\,,\nonumber \\
g_{tj}&=-4 U_j+\mathcal{O}(v^5)\,,\nonumber \\
g_{jk}&=\left(1-2U\right) \delta_{jk}+\mathcal{O}(v^4)\,,
\end{align}
where the different PN potentials, as well as time derivatives, can be organized in powers of a characteristic velocity $v\ll1$:
\begin{equation}
 U=\mathcal{O}(v^2),~
 U_j=\mathcal{O}(v^3),~
 \psi=\mathcal{O}(v^4),~
 \partial_t\sim v\,\partial_j.
 \label{eq:PN_counting}
\end{equation}
No independent PN order needs to be assigned to $\hat\Phi$. The scaling of its density follows directly from the Poisson equation. 
The Ricci-tensor components for the 1PN metric are:
\begin{align}
R_{tt}=& \nabla^2U- \partial_t^2 U+4 U \nabla^2 U+ \nabla^2 \psi+ \mathcal{O}(v^6)\,, \nonumber \\
R_{tj}=&2 \nabla^2U_j+ \mathcal{O}(v^5)\,, \nonumber \\
R_{jk}=&\nabla^2U\delta_{jk}+\mathcal{O}(v^4)\,,
\end{align}
and the harmonic coordinate condition gives the constraint:
\begin{equation}
	\partial_tU-\partial_j U^j=0\,.
\end{equation}
For a cloud in a single scalar eigenmode, the stress-energy tensor is stationary and axisymmetric, so all metric functions share these symmetries and $\partial_tU=0$. We further decompose the PN potentials according to their physical sources,
\begin{equation}
 U=U^{\rm BH}+U^S,\qquad
 U_j=U_j^{\rm BH}+U_j^S,\qquad
 \psi=\psi^{\rm BH}+\psi^S.
 \label{eq:potential_decomposition}
\end{equation}
The superscripts ${\rm BH}$ and $S$ label the BH and cloud contributions, respectively. The cloud potentials are evaluated using the prescribed hydrogenic profile and eigenfrequency, retaining the nonlinear source terms present in the 1PN equations. The resulting potentials include the cloud's self-gravity, but its feedback on the scalar profile and eigenfrequency is neglected.
\begin{figure*}
    \centering
    \includegraphics[width=\textwidth]{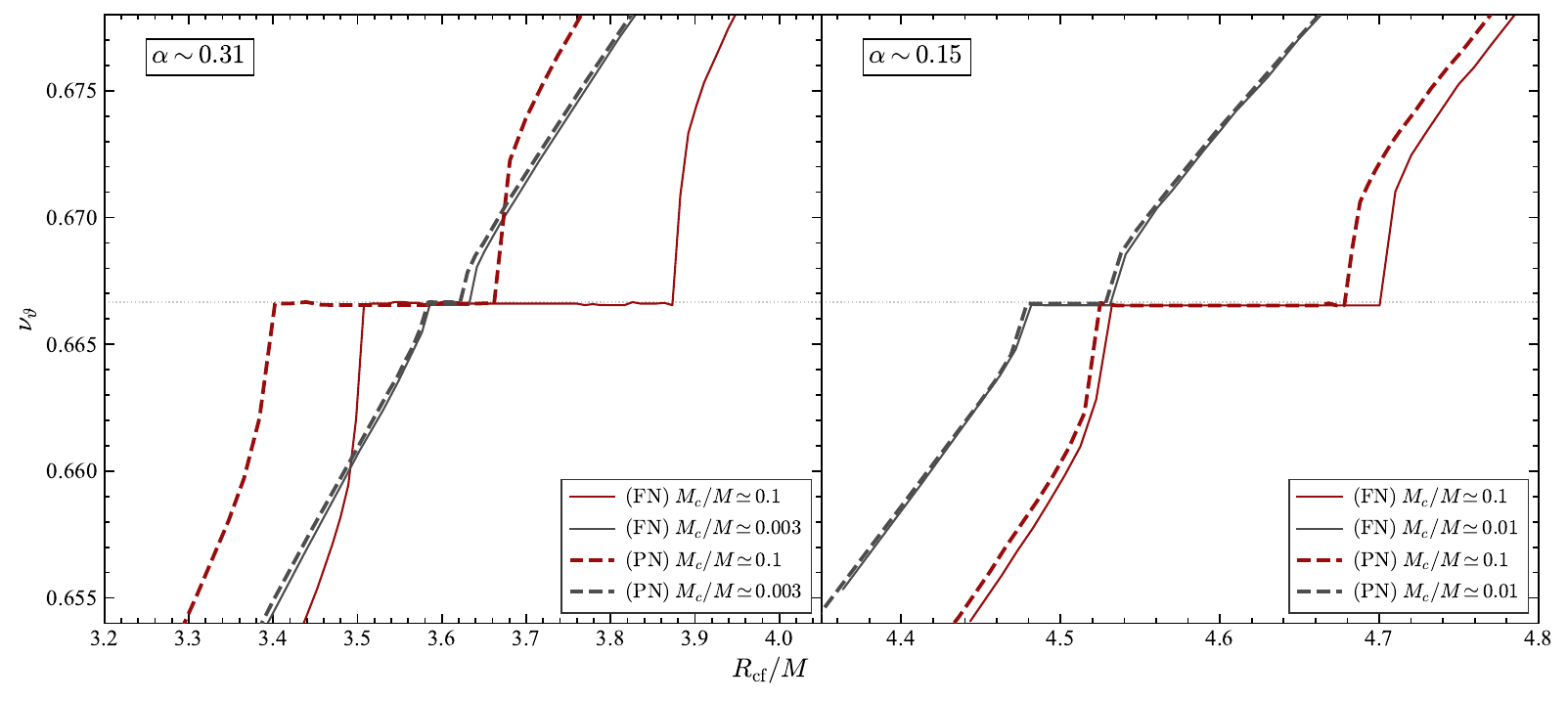}
    \caption{Rotation curves around the region of the $2/3$ resonance obtained with the PN (dashed lines) and FN (solid lines) geometries, expressed in terms of the circumferential radius $R_{\rm cf}$. The \emph{left panel} corresponds to $\alpha\simeq0.31$, $E/m_p=0.95$, and $L_z/m_p=2.6M$; the \emph{right panel} corresponds to $\alpha\simeq0.15$, $E/m_p=0.98$, and $L_z/m_p=3.3M$.}
    \label{fig:FN_vs_PN}
\end{figure*}

The relevant components of the scalar trace-reversed
stress-energy tensor are
\begin{align}
 \widetilde T_{tt}^S
 &=\mu^2|\hat\Phi|^2
 +4\mu\,{\rm Im}\left(\hat\Phi\,\partial_t\hat\Phi^*\right)
 -2\mu^2U|\hat\Phi|^2+\mathcal{O}(v^6),\nonumber\\
 \widetilde T_{tj}^S
 &=2\mu\,{\rm Im}\left(\hat\Phi\,\partial_j\hat\Phi^*\right)
 +\mathcal{O}(v^5),\nonumber\\
 \widetilde T_{jk}^S
 &=\mu^2|\hat\Phi|^2\delta_{jk}+\mathcal{O}(v^4).
\end{align}
Combining these expressions with the Ricci tensor gives a set of Poisson
equations for the cloud-sourced potentials:
\begin{subequations}
\label{eq:PN_Poisson_eqs}
\begin{align}
\nabla^2U^S&=8\pi \mu^2|\hat{\Phi}|^2\,,\label{subeq:scalar_potential}\\
\nabla^2 U^S_j&=8 \pi\mu \,{\rm Im}\left(\hat{\Phi}\,\partial_j \hat{\Phi}^*\right)\,,\label{subeq:vector potential}\\
\nabla^2\psi^S&=16\pi\mu \Big[2{\rm Im}\left(\hat{\Phi}\, \partial_t \hat{\Phi}^*\right)-3\mu U |\hat{\Phi}|^2\Big]\,.\label{subeq:scalar_1PN_correction}
\end{align}
\end{subequations}
These equations have the same form as those obtained in the context of Newtonian boson stars~\cite{Annulli:2020lyc}, although here the scalar profile is an hydrogenic state supported by the central BH.

To our knowledge, this is the first explicit derivation of a 1PN metric describing scalar clouds, although parts of it were previously derived in Refs.~\cite{Ferreira:2017pth,Boskovic:2026eth}. The derivations and the closed-form expressions for all the metric potentials for the $\{n,\ell,m\}=\{0,1,1\}$ mode are presented in App.~\ref{app:pn_potentials}. These results provide the analytical PN geometry that we use below and may also be useful for future studies of orbital dynamics and GW signatures in BH--cloud spacetimes.

\subsection{Resonant islands: comparison between PN and FN solutions}
\label{sec:pn_fr_comparison}
In this section, we compare results for rotation-number curves obtained with the PN and FN geometries after matching their BH parameters and gravitational cloud masses. For the FN geometries, the central BH mass and spin are calculated from the behavior at the horizon as is described in Ref. \cite{May:2024npn}. For each FN configuration, the PN background is assigned the same BH mass and spin. The normalization of the hydrogenic scalar profile is then chosen so that the asymptotic cloud mass of the PN metric equals the FN Komar cloud mass, following App.~\ref{app:cloud_mass_matching}. Thus, the cloud mass choices quoted below refer to the matched gravitational mass. We should also note that the ``PN'' metric used in the geodesic calculations is $g_{\mu\nu}^{\rm PN}=g_{\mu\nu}^{\rm Kerr}+\delta g_{\mu\nu}^{S}$, with the Kerr background retained in full and only the cloud-induced correction treated through 1PN order. The correction $\delta g_{\mu\nu}^{S}$ is obtained from Eq.~\eqref{eq:1pNmetric} by subtracting its vacuum contribution with the same BH parameters.

Since the PN and FN metrics use different radial coordinates, we cannot perform a direct comparison at the same coordinate value of $r$. Therefore, to compare results in both metrics, we express all calculations in terms of the equatorial circumferential radius,
\begin{equation}
    R_{\rm cf}
    =\frac{1}{2\pi}\int_0^{2\pi}
    d\varphi\,
    \sqrt{g_{\varphi\varphi}(r,\theta=\pi/2)}.
    \label{eq:circumferential_radius}
\end{equation}
The quantity $2\pi R_{\rm cf}$ measures the proper circumference of the orbit generated by the axial Killing vector on the equatorial plane. It therefore provides a common geometric radial measure for the PN and FN spacetimes. 

Fig.~\ref{fig:FN_vs_PN} compares the resulting rotation curves as a function of $R_{\rm cf}$, with solid and dashed lines denoting the FN and PN descriptions, respectively. Red and black curves correspond to the more and less massive clouds, respectively. In all cases we see a finite plateau associated with the $\nu_\vartheta=2/3$ resonance. The intervals and widths quoted below describe the sampled flat portions of the curves. Their endpoints are approximate and are limited by the radial sampling and finite integration time.

The right panel corresponds to $\alpha\simeq0.15$, $E/m_p=0.98$, and $L_z/m_p=3.3M$. For the less massive cloud, $M_c/M\simeq0.01$, the curves obtained with the PN and FN metrics very closely track one another, with both predicting plateaus lying approximately within $4.48\lesssim R_{\rm cf}/M\lesssim4.53$ and having width $\Delta R_{\rm cf}\simeq0.05M$. The agreement remains reasonable even for the more massive clouds with $M_c/M\simeq0.1$, although the plateau in the PN metric is slightly narrower and shifted toward smaller $R_{\mathrm{cf}}$ relative to the one in the FN metric.
In this case, the plateau in the PN metric lies in the approximate range $4.52\lesssim R_{\rm cf}/M\lesssim4.68$, corresponding to a width $\Delta R_{\rm cf}\simeq 0.16M$, whereas for the FN solution it  spans the range $4.53\lesssim R_{\rm cf}/M\lesssim4.70$, i.e., a width of $\Delta R_{\rm cf}\simeq 0.17M$. 
The non-resonant portions of the rotation curves also follow similar trends. Therefore, for $\alpha\sim 0.15$, the PN description reproduces both the resonant structure and its approximate radial extent to a very good extent.

The left panel corresponds to results for BH--cloud configurations with $\alpha\simeq0.31$, $E/m_p=0.95$, and $L_z/m_p=2.6M$. For the light cloud, with $M_c/M\simeq0.003$, the two curves remain close, but their plateau edges are not exactly identical. The sampled PN plateau spans approximately $3.58\lesssim R_{\rm cf}/M\lesssim3.62$, whereas the FN plateau extends to $3.58\lesssim R_{\rm cf}/M\lesssim3.63$. Their corresponding widths are $\sim0.04M$ and $\sim0.05M$, respectively. This difference is comparable to one radial sampling interval. This case shows that even for $\alpha\simeq0.31$, where one could expect the PN approximation to fail, the PN metric still provides reasonably accurate results, as long as the mass of the cloud is sufficiently small. A substantially larger discrepancy appears for $M_c/M\simeq0.1$ with the same coupling $\alpha$. The PN plateau in this case spans approximately $3.40\lesssim R_{\rm cf}/M\lesssim3.66$, whereas the FN plateau lies within $3.51\lesssim R_{\rm cf}/M\lesssim3.87$. The corresponding sampled widths are $\sim0.26M$ and $\sim0.36M$. The PN model, thus, starts failing for large cloud masses and couplings, i.e. for more compact configurations, since it underestimates the plateau width, by at least $\mathcal{O}(10^{-1})$ with respect to the plateau that results from the FN configuration.

Finally, there is an additional subtlety in comparing configurations with different cloud masses worth noticing. In the PN construction, the cloud mass can in principle be varied independently of the background BH parameters. In a self-consistent FN family, by contrast, the cloud mass, scalar frequency, horizon position, and BH angular momentum are all related through the nonlinear field equations and the synchronization condition \cite{Herdeiro:2014goa,May:2024npn}, so changing $M_c$ necessarily changes the BH parameters as well. We therefore match each PN--FN pair separately, and the comparison across cloud masses in Fig.~\ref{fig:FN_vs_PN} should not be interpreted as varying $M_c$ at fixed BH parameters.

\section{Discussion and Conclusions}\label{Sec:Discussion and Conclusions}

Orbital resonances are intrinsic to Kerr EMRIs, but in the conservative, integrable Kerr geometry they lie on zero-width tori, where the libration zone shrinks to a typical stable point. We have shown, for the first time, that a scalar cloud around a spinning BH, qualitatively changes this phase-space structure. The cloud-induced deformation breaks geodesic integrability and replaces resonant tori with Birkhoff chains that possess finite-width nested resonant islands, which appear as plateaus in the corresponding rotation curves. We identified the dominant $2/3$ resonance in both numerically computed spacetimes, as well as in an approximate PN description of the cloud's perturbation on the geometry. We found resonant islands even in configurations in which the cloud contains only a small fraction of the BH mass, and showed that the width of the resonant islands increases with the compactness of the cloud. Scalar clouds therefore provide a physically-motivated environmental source of non-integrability around rotating BHs. Together with previous studies of other environments, such as dark matter halos and tidal perturbations~\cite{Destounis:2025tjn,Destounis:2026tet}, our results suggest that astrophysical environments generically break the integrability of geodesic motion around BHs.

We also showed that for a cloud as massive as the BH, $M_c \sim M$, the phase space exhibits a richer structure: $2/3$, $4/5$, $6/7$, and $6/8$
resonant chains coexist with regular KAM tori and with chaotic layers surrounding unstable periodic points. This finding highlights the distinction between non-integrability and chaos. Finite-width resonant
islands already establish non-integrability for weak clouds, whereas clear chaotic layers appear only in the comparable-mass configuration. A broader survey of cloud and orbital parameters is needed to determine
how the phase-space structure evolves and whether comparably prominent chaotic features can arise for weak clouds in some regions of parameter space.

Finally, we compared results obtained with the fully analytical PN metric against the ones obtained with FN solutions. 
The two approaches yield closely agreeing rotation curves and $2/3$-resonance widths for weak clouds at both $\alpha\simeq0.15$ and $\alpha\simeq0.31$. Appreciable differences arise, however, when the cloud is sufficiently massive and compact, namely for a configuration with $M_c/M\simeq0.1$ and $\alpha\simeq0.31$. Thus, the PN construction captures the onset of non-integrability and the qualitative resonant structure for weak clouds, but the FN solutions are required for reliable quantitative predictions when the cloud is compact and carries a substantial fraction of the mass.

All effects identified here are conservative environmental effects arising from geodesic motion in stationary BH--cloud spacetimes. Radiation reaction, self-force corrections, and the dynamical response of the cloud due to the secondary are deliberately not included. Consequently, the present calculation determines the conservative resonant structure but not the time the secondary spends in a resonance or the resulting accumulated gravitational waveform dephasing. Predicting these observables requires evolving the orbit through the
resonant islands while accounting for gravitational and scalar torques~\cite{Destounis:2021mqv,Destounis:2021rko,Dyson:2025dlj}. The analytical PN metric derived here offers a tractable BH--cloud background for this purpose. It can be used both to study conservative environmental effects more broadly and as a starting point for incorporating dissipative dynamics and constructing waveform models. This is a natural direction for future work.

\begin{acknowledgments}
QX acknowledges support from FCT through Grant No.~\href{https://doi.org/10.54499/2025.01546.BD}{2025.01546.BD}.
KD acknowledges financial support provided by FCT–Fundação para a Ciência e a Tecnologia, I.P., under the Scientific Employment Stimulus – Individual Call – Grant No. \href{https://sciproj.ptcris.pt/12680EEC}{2023.07417.CEECIND/CP2830/CT0008}.
RB acknowledges financial support from FCT through the ERC-Portugal programme project ``GravNewFields.'' 
YG acknowledges financial support from the National Natural Science Foundation of China (Grants Nos. 124B2098, 12447105, 12075136) and from the Natural Science Foundation of Shandong Province (Grant No. ZR2020MA094). 
TM acknowledges funding from the European Union's Horizon 2020 research and innovation programme under the Marie Skłodowska-Curie grant agreement No.~101007855, and support from the Simons Foundation International and the Simons Foundation through Simons Foundation grant SFI-MPS-BH-00012593-11.
KD and TM also acknowledge funding from the European Union's Horizon Europe research and innovation programme under the Marie Skłodowska-Curie Actions project ``EinsteinWaves,'' Grant Agreement No.~101131233.
We acknowledge financial support from FCT for the Center for Astrophysics and Gravitation (CENTRA/IST/ULisboa) through Grants No.~\href{https://doi.org/10.54499/UID/PRR/00099/2025}{UID/PRR/00099/2025} and No.~\href{https://doi.org/10.54499/UID/00099/2025}{UID/00099/2025}.
\end{acknowledgments}

\appendix
\section{Fully nonlinear solutions: Configurations and convergence}\label{app:geodesic_convergence}
\subsection{Configurations}

The resolutions used for the FN solutions of this paper have $391$ radial points and $46$ angular points, i.e., $(391, 46)$. We have also used two lower resolutions with $(261, 35)$ and $(167, 20)$ to assess the behavior of the results with respect to the resolution of the background metric. The consecutive resolutions differ by approximately a factor of $2/3$.

The configurations used in the main calculations are listed in Table~\ref{tab:configurations}.
\begin{table}[h]
\centering
\caption{FN configurations used in the geodesic analysis.}
\label{tab:configurations}
\setlength{\tabcolsep}{6pt}
\begin{ruledtabular}
\begin{tabular}{ccc}
$\alpha$ & $M_c/M$ & $\omega/\mu$ \\
\hline
$0.1492$ & $0.01191$  & $0.9971$ \\
$0.1408$ & $0.09451$  & $0.9971$ \\
\hline
$0.3168$ & $0.00308$ & $0.9860$ \\
$0.3042$ & $0.01278$  & $0.9870$ \\
$0.3102$ & $0.03268$  & $0.9860$ \\
$0.3106$ & $0.09814$  & $0.9845$ \\
$0.3062$ & $0.28566$  & $0.9755$ \\
$0.3103$ & $1.00805$   & $0.9540$ \\
\end{tabular}
\end{ruledtabular}
\end{table}

\subsection{Convergence test}
To check convergence of our results, we focus on the most prominent stable resonant island, i.e. the $2/3$-resonant island. We have found that the $2/3$ island widths calculated with all three resolutions stated above for the FN metrics are identical up to the precision of the island-finder. This means that the difference between them is $< 2\times10^{-2}\%$. We also test the stability of the island width with an internal parameter of the geodesic solver which specifies how many effective digits of accuracy should be sought in the final result. We find that increasing that parameter in increments, from 8 to 12 digits of internal accuracy, leads to island widths that are again identical up to the precision of the island-finder, with a difference $< 2\times10^{-2}\%$. We know that the metric solutions themselves converge \cite{May:2024npn}, and the geodesic solver is a published \texttt{Mathematica} method with established convergence. We conclude that the dominant source of error in our calculation is the precision of the island-finder, with an error of $\mathcal{O}(10^{-3})M$.

\section{Post-Newtonian potentials sourced by the scalar cloud}
\label{app:pn_potentials}

We derive here the scalar-cloud contribution to the 1PN metric used in the main text.  In harmonic coordinates, the PN potentials satisfy Eq.~\eqref{eq:PN_Poisson_eqs}.
These equations separate the cloud correction into a Newtonian scalar potential, a gravitomagnetic vector potential sourced by the mass current, and a scalar 1PN correction.  We solve them in this
order below. 
We impose regularity at the origin and asymptotic flatness, and specialize to the most interesting hydrogenic $|211\rangle$ state ($\bar{n}=2, \ell=1, m=1$).  Throughout this
appendix, we define dimensionless quantities $x\equiv\mu\alpha r$ for convenience.

\subsection{Newtonian self-gravity}

The $|211\rangle$ scalar field has dipolar angular dependence
($\ell=1$), but the Newtonian potential is sourced by the density $\propto|\hat{\Phi}|^2$ [see Eq.~\eqref{subeq:scalar_potential}].  The harmonic time-dependence and azimuthal phase cancel in the density, so the resulting potential is stationary and axisymmetric.  Moreover, the product of two $\ell=1$ harmonics contains only the even-parity monopolar and
quadrupolar sectors for this state.  We therefore begin by decomposing
the Newtonian potential and its source using spherical harmonics
\begin{align}
 U^S(r,\theta)&=\sum_L U_L(r)Y_{L0}(\theta),
 \nonumber\\
 |\hat{\Phi}(r,\theta)|^2
 &=\sum_L\rho_L(r)Y_{L0}(\theta).
 \label{eq:app_scalar_expansion}
\end{align}
For a radial source $S_L=8\pi\mu^2\rho_L$, Eq.~\eqref{subeq:scalar_potential} becomes
\begin{equation}
 \frac{1}{r^2}\frac{d}{dr}
 \left(r^2\frac{dU_L}{dr}\right)
 -\frac{L(L+1)}{r^2}U_L=S_L.
 \label{eq:app_radial_poisson}
\end{equation}
The solution satisfying the stated boundary conditions is
\begin{equation}
\begin{aligned}
 U_L(r)=-\frac{1}{2L+1}\Bigg[
 &\frac{1}{r^{L+1}}
 \int_0^r s^{L+2}S_L(s)\,ds
 \\
 &+r^L\int_r^\infty
 s^{1-L}S_L(s)\,ds\Bigg].
\end{aligned}
\label{eq:app_green_function}
\end{equation}
This equation will also be used below to obtain the other PN potentials.

For the $|211\rangle$ state, the hidrogenic radial wave function reduces to
\begin{equation}
 R^h_{0,1}(r)=
 \frac{\sqrt{M_c}}{\sqrt{2}\mu}
 \frac{(\mu\alpha)^{3/2}}{\sqrt{24}}\,
 x e^{-x/2},
 \label{eq:app_radial_wavefunction}
\end{equation}
and hence
\begin{equation}
 \mathcal{P}(r)\equiv
 |R^h_{0,1}(r)|^2
 =\frac{M_c}{2\mu^2}
 \frac{(\mu\alpha)^5}{24}r^2e^{-x}.
 \label{eq:app_radial_density}
\end{equation}
The angular density,
\begin{equation}
 |Y_{11}|^2=
 \frac{1}{\sqrt{4\pi}}Y_{00}
 -\frac{1}{2\sqrt{5\pi}}Y_{20},
 \label{eq:app_angular_density}
\end{equation}
makes the selection rule explicit: the density is axisymmetric and
contains only $L=0$ and $L=2$ spherical harmonics. Therefore,
\begin{equation}
 \rho_0(r)=\frac{\mathcal{P}(r)}{\sqrt{4\pi}},
 \qquad
 \rho_2(r)=-\frac{\mathcal{P}(r)}{2\sqrt{5\pi}}.
 \label{eq:density_coefficients_211}
\end{equation}
These two coefficients exhaust the multipolar content of the
Newtonian source.  Since their radial dependence is a polynomial
times $e^{-x}$, the integrals in Eq.~\eqref{eq:app_green_function} can be evaluated in
closed form.  Substitution into Eq.~\eqref{eq:app_green_function}
yields the
monopolar contribution
\begin{equation}
\begin{aligned}
 U_0(r)={}&-\frac{\sqrt{4\pi}M_c}{r}
 \Bigg[1-e^{-x}\Bigg(
 1+\frac{3x}{4}
 +\frac{x^2}{4}+\frac{x^3}{24}
 \Bigg)\Bigg],
\end{aligned}
\label{eq:app_U0}
\end{equation}
and the quadrupolar contribution
\begin{equation}
\begin{aligned}
 U_2(r)={}&
 \frac{\sqrt{\pi/5}M_c}
 {12\mu^2\alpha^2r^3}
 \Bigg[
 144-e^{-x}\left(
 144\sum_{k=0}^{4}\frac{x^k}{k!}
 +x^5\right)\Bigg].
\end{aligned}
\label{eq:app_U2}
\end{equation}
The cloud potential is thus
\begin{equation}
 U^S(r,\theta)=
 U_0(r)Y_{00}(\theta)
 +U_2(r)Y_{20}(\theta).
 \label{eq:U_cloud_multipoles}
\end{equation}
In particular, its monopole approaches $-M_c/r$ at large radii,
whereas the quadrupole falls off as $r^{-3}$, as expected.

\subsection{Gravitomagnetic potential}

The vector potential is sourced by the scalar mass current rather
than by the density itself [see Eq.~\eqref{subeq:vector potential}].  For a mode of the form
$\hat{\Phi}=R(r)\Theta(\theta)e^{im\phi-iE_{n \ell}t}$, the
radial and polar derivatives do not contribute to the imaginary part
of $\hat{\Phi}\nabla\hat{\Phi}^*$ because $R$ and $\Theta$ may be
chosen real.  The derivative of the azimuthal phase is therefore the
only surviving contribution, and the current is purely azimuthal.
In an orthonormal spherical basis its nonvanishing component is
\begin{equation}
\begin{aligned}
 S_{\hat{\phi}}(r,\theta)
 &=
 8\pi\mu\,{\rm Im}
 \left(\hat{\Phi}\nabla\hat{\Phi}^*\right)_{\hat{\phi}}
 \\
 &=-8\pi\mu
 \frac{m}{r\sin\theta}|\hat{\Phi}|^2.
\end{aligned}
\label{eq:source_phi_hat}
\end{equation}
An axisymmetric azimuthal vector field obeys
\begin{equation}
 \left(\nabla^2\boldsymbol{U}^S\right)_{\hat{\phi}}
 =
 \nabla^2 U_{\hat{\phi}}^S
 -\frac{U_{\hat{\phi}}^S}
 {r^2\sin^2\theta}.
 \label{eq:app_vector_laplacian}
\end{equation}
The additional angular term in this equation is the order-one associated
Legendre operator.  The appropriate angular basis is therefore
$P_L^1(\cos\theta)$:
\begin{align}
 U_{\hat{\phi}}^S(r,\theta)
 &=\sum_{L=1}^{\infty}
 u_L(r)P_L^1(\cos\theta),
 \nonumber\\
 S_{\hat{\phi}}(r,\theta)
 &=\sum_{L=1}^{\infty}
 s_L(r)P_L^1(\cos\theta).
 \label{eq:app_vector_expansion}
\end{align}
The radial equation for $(u_L,s_L)$ is identical to
Eq.~\eqref{eq:app_radial_poisson}, so the same solution applies.

For the $|211\rangle$ state, $m=1$ and
$|\hat{\Phi}|^2/\sin\theta\propto\sin\theta$.  Since
$P_1^1(\cos\theta)=-\sin\theta$, this angular dependence projects
entirely onto the $L=1$ vector harmonic.  The source consequently
contains only the $L=1$ sector,
\begin{equation}
\begin{aligned}
 s_1(r)
 &=\frac{3\mu}{r}\mathcal{P}(r)=\frac{M_c}{\mu}
 \frac{(\mu\alpha)^5}{16}
 r e^{-x}.
\end{aligned}
\label{eq:app_s1}
\end{equation}
Equation~\eqref{eq:app_green_function} then gives
\begin{equation}
\begin{aligned}
 u_1(r)={}&-\frac{M_c}{2\mu r^2}
 \Bigg[
 1-e^{-x}\Bigg(
 1+x
 +\frac{x^2}{2}+\frac{x^3}{8}
 \Bigg)\Bigg].
\end{aligned}
\label{eq:app_u1}
\end{equation}
Having solved the only nonvanishing radial mode, we reconstruct the
vector potential and convert from the orthonormal component to the
covariant component entering the 1PN metric.  The latter follows from
$U_\phi^S=r\sin\theta\,U_{\hat{\phi}}^S$:
\begin{equation}
\begin{aligned}
 U_\phi^S(r,\theta)
 ={}&\frac{M_c}{2\mu r}
 \Bigg[
 1-e^{-x}\Bigg(
 1+x+\frac{x^2}{2}+\frac{x^3}{8}
 \Bigg)\Bigg]\sin^2\theta.
\end{aligned}
\label{eq:app_Uphi}
\end{equation}
This expression is regular on the symmetry axis.  At large radii,
$U_\phi^S\to M_c\sin^2\theta/(2\mu r)$.  Since the $|211\rangle$
cloud carries angular momentum $J_c=mM_c/\mu=M_c/\mu$, the
corresponding metric component satisfies
$g_{t\phi}\to-2J_c\sin^2\theta/r$, in agreement with the weak-field
Lense--Thirring limit.

\subsection{Post-Newtonian trace potential}
Equation~\eqref{subeq:scalar_1PN_correction} contains two physically distinct sources: the nonrelativistic bound-state
energy and the gravitational interaction of the cloud with the
total Newtonian potential.  The latter further separates according
to whether the cloud density couples to the central black-hole
potential or to its own Newtonian potential.  We accordingly write
\begin{equation}
 \psi^S=
 \psi_E^S+
 \psi_{{\rm BH,int}}^S+
 \psi_{{\rm S,int}}^S.
 \label{eq:psi_total_decomposition}
\end{equation}

\subsubsection{Bound-state energy}

For $\hat{\Phi}\propto e^{-iE_{n \ell}t}$,
$\partial_t\hat{\Phi}^*=iE_{n \ell}\hat{\Phi}^*$, and the energy source is
$32\pi\mu E_{n \ell}|\hat{\Phi}|^2$.  Comparing it with the Poisson equation
for $U^S$ gives, by linearity,
\begin{equation}
 \psi_E^S=\frac{4E_{n \ell}}{\mu}U^S.
 \label{eq:psi_energy}
\end{equation}
Thus, the energy correction has exactly the same monopolar and
quadrupolar content as $U^S$; no additional radial integration is
required.

\subsubsection{BH--cloud interaction}

At Newtonian order, $U^{\rm BH}=-M/r$.  The corresponding source,
\begin{equation}
 S_{{\rm BH,int}}
 =48\pi\mu^2\frac{M}{r}
 |\hat{\Phi}|^2,
 \label{eq:S_BH_int}
\end{equation}
inherits the monopolar and quadrupolar structure of the cloud
density.  Indeed, multiplication by the spherically symmetric
black-hole potential cannot mix angular multipoles.  Its radial
components are
\begin{align}
 \mathcal{S}^{\rm BH}_0(r)
 &=
 \frac{\sqrt{\pi}}{2}
 MM_c(\mu\alpha)^5
 r e^{-x},
 \label{eq:S_BH_0}
 \\
 \mathcal{S}^{\rm BH}_2(r)
 &=
 -\frac{1}{2}
 \sqrt{\frac{\pi}{5}}\,
 MM_c(\mu\alpha)^5
 r e^{-x}.
 \label{eq:S_BH_2}
\end{align}
The interaction potential therefore has the form
\begin{equation}
 \psi_{{\rm BH,int}}^S
 =
 \psi_0^{{\rm BH,int}}Y_{00}
 +\psi_2^{{\rm BH,int}}Y_{20}.
 \label{eq:psi_BH_decomposition}
\end{equation}
Using a solution of the form given in Eq.~\eqref{eq:app_green_function}, with the corresponding sources, gives
\begin{equation}
\begin{aligned}
 \psi_0^{{\rm BH,int}}(r)
 ={}&
 \frac{\sqrt{\pi}MM_c\mu\alpha}{2r}
 \left[
 e^{-x}\left(x^2+4x+6\right)-6
 \right],
\end{aligned}
\label{eq:psi_BH_int_0}
\end{equation}
and
\begin{equation}
\begin{aligned}
 \psi_2^{{\rm BH,int}}(r)
 &={}
 \frac{\sqrt{5\pi}MM_c}
 {10\mu\alpha r^3}
 \Big[
 24
 \\
 &-e^{-x}\big(
 x^4+4x^3+12x^2
 +24x+24
 \big)\Big].
\end{aligned}
\label{eq:psi_BH_int_2}
\end{equation}
No higher multipoles are generated because the black-hole
potential $U^{\rm BH}$ is spherically symmetric.  The leading asymptotic term is
\begin{equation}
 \psi_{{\rm BH,int}}^S(r,\theta)
 =
 -\frac{3}{2}\frac{\alpha^2M_c}{r}
 +\mathcal{O}(r^{-3}).
 \label{eq:psi_BH_asymptotic}
\end{equation}

\subsubsection{Cloud self-interaction}

The remaining interaction accounts for the cloud's coupling to its
own Newtonian potential and is sourced by
\begin{equation}
 S_{{\rm S,int}}
 =
 -48\pi\mu^2
 U^S|\hat{\Phi}|^2.
 \label{eq:S_self_int}
\end{equation}
Since $U^S$ is itself proportional to the cloud normalization, this
source is quadratic in the cloud mass, in contrast to the
BH--cloud contribution above.  Both factors contain $L=0$ and $L=2$
components.  Their product therefore generates $L=0,2,4$, as follows
from
\begin{align}
 Y_{00}Y_{00}
 &=
 \frac{1}{\sqrt{4\pi}}Y_{00},
 \nonumber\\
 Y_{00}Y_{20}
 &=
 \frac{1}{\sqrt{4\pi}}Y_{20},
 \nonumber\\
 Y_{20}Y_{20}
 &=
 \frac{1}{\sqrt{4\pi}}Y_{00}
 +\frac{\sqrt{5}}{7\sqrt{\pi}}Y_{20}
 \nonumber 
 +\frac{3}{7\sqrt{\pi}}Y_{40}.
 \label{eq:Y20_squared}
\end{align}
This angular-product decomposition explains the appearance of the
hexadecapole even though neither the Newtonian density nor $U^S$
contains an $L=4$ component.  Using
Eqs.~\eqref{eq:density_coefficients_211} and
\eqref{eq:U_cloud_multipoles}, the radial sources become
\begin{equation}
\begin{aligned}
 \mathcal{S}^{\rm S}_0(r)
 =
 -48\pi\mu^2
 \frac{1}{\sqrt{4\pi}}
 \left(
 U_0\rho_0+U_2\rho_2
 \right),
\end{aligned}
\label{eq:S_self_0}
\end{equation}
\begin{equation}
\begin{aligned}
 \mathcal{S}^{\rm S}_2(r)
 =
 -48\pi\mu^2
 \Bigg[
 &\frac{U_0\rho_2+U_2\rho_0}
 {\sqrt{4\pi}}+\frac{\sqrt{5}}{7\sqrt{\pi}}
 U_2\rho_2
 \Bigg],
\end{aligned}
\label{eq:S_self_2}
\end{equation}
and
\begin{equation}
 \mathcal{S}^{\rm S}_4(r)
 =
 -48\pi\mu^2
 \frac{3}{7\sqrt{\pi}}
 U_2\rho_2.
 \label{eq:S_self_4}
\end{equation}
Consequently,
\begin{equation}
\begin{aligned}
 \psi_{{\rm S,int}}^S
 ={}&
 \psi_0^{{\rm S,int}}Y_{00}
 +\psi_2^{{\rm S,int}}Y_{20} +\psi_4^{{\rm S,int}}Y_{40}.
\end{aligned}
\label{eq:psi_self_decomposition}
\end{equation}

Applying Eq.~\eqref{eq:app_green_function} independently to the
$L=0$, $L=2$, and $L=4$ sources completes the self-interaction
solution.  The resulting integrals can be evaluated in closed form.
The monopolar component is
\begin{equation}
\begin{aligned}
&\psi^{{\rm S,int}}_0(r) = \\
    &\frac{\sqrt{\pi}\alpha \mu {M_c}^2}{320 r} \Bigg[ -711 + 32 e^{-x} \left(36 + 20x + 5x^2 \right)- e^{-2x} \\
    & \times \Big( 441 + 566x + 346x^2 + 116x^3 + 22x^4 + 2x^5 \Big) \Bigg],
\end{aligned}
\label{eq:psi_self_0}
\end{equation}
The quadrupolar component is
\begin{equation}
\begin{aligned}
    &\psi^{{\rm S,int}}_2(r) = \\
    &\frac{\sqrt{\pi}{M_c}^2}{448 \sqrt{5} \mu\alpha r^3} \Bigg\{ 
    5049 - 192 x^5 \Big[ \Gamma(0, x) - \Gamma(0, 2x) \Big] \\
    &\quad - 32 e^{-x} \left( 204 + 204x + 102x^2 + 34x^3 + x^4 \right) \\
    &\quad + e^{-2x} \bigg( 1479 + 2958x + 2958x^2 + 1972x^3 \\
    &\quad + 746x^4 + 244x^5 + 44x^6 + 4x^7 \bigg) \Bigg\},
\end{aligned}
\label{eq:psi_self_2}
\end{equation}
while the hexadecapolar component is
\begin{equation}
\begin{aligned}
    &\psi^{{\rm S,int}}_4(r) =\\
    &\frac{\sqrt{\pi}{M_c}^2}{6720  \alpha^3 \mu^3 r^5} \Bigg\{ 
    -23985 + 64 x^9 \Big[ \Gamma(0, x) - \Gamma(0, 2x) \Big] \\
    &\quad + 64 e^{-x} \Big( 720 + 720x + 360x^2 + 120x^3 \\
    &\quad + 30x^4 + 6x^5 - 2x^6 + x^7 - x^8 \Big) \\
    &\quad - e^{-2x} \Big( 22095 + 44190x + 44190x^2 + 29460x^3 \\
    &\quad + 14730x^4 + 5892x^5 + 1772x^6 + 424x^7 + 34x^8 + 6x^9 \Big) \Bigg\}.
\end{aligned}
\label{eq:psi_self_4}
\end{equation}
Here $\Gamma(0,x)$ is the upper incomplete gamma function.  The
$L=4$ sector is generated exclusively by the product of the
quadrupolar cloud density and the quadrupolar Newtonian potential;
the spherically symmetric BH potential cannot produce this
component.  Combining the bound-state contribution with the
BH--cloud and cloud self-interaction terms according to
Eq.~\eqref{eq:psi_total_decomposition} completes $\psi^S$.
Together with Eqs.~\eqref{eq:app_U0}, \eqref{eq:app_U2}, and
\eqref{eq:app_Uphi}, this provides all scalar-cloud contributions
required to construct the 1PN metric. In particular, inserting
$U=U^{\rm BH}+U^S$ into the nonlinear term in
Eq.~\eqref{eq:1pNmetric} gives
\begin{equation}
 U^2=(U^{\rm BH})^2+2U^{\rm BH}U^S+(U^S)^2.
 \label{eq:U_squared_decomposition}
\end{equation}
The three terms scale respectively as $M^2$, $MM_c$, and $M_c^2$,
but all contribute to $g_{tt}$ at $\mathcal{O}(v^4)$.  This completes
the cloud-induced 1PN metric.

\section{Cloud mass matching between PN and fully nonlinear metric}
\label{app:cloud_mass_matching}

Comparing the PN and FN geometries requires distinguishing the normalization of the hydrogenic scalar profile from the gravitational mass attributed to the cloud.  We denote the former by
$M_c^{\rm norm}$,
\begin{equation}
 2\mu^2\int d^3x\,|\hat{\Phi}|^2
 =M_c^{\rm norm}.
 \label{eq:Mc_normalization}
\end{equation}
This is the parameter denoted by $M_c$ in the hydrogenic wave function and in the potentials of App.~\ref{app:pn_potentials}. It agrees with the cloud's gravitational mass at Newtonian order, but the two quantities differ once the 1PN contributions are retained.

For the FN solutions, the scalar contribution to the mass is defined by the Komar volume integral
\begin{equation}
 M_c^{\rm K}
 =-\int_{\Sigma}dS_a
 \left(2T^a{}_b t^b-Tt^a\right),
 \label{eq:Mc_Komar}
\end{equation}
where $\Sigma$ extends from the horizon to spatial infinity, $t^a$ is the stationary Killing field normalized at infinity.
The FN mass decomposition is $M_{\rm ADM}^{\rm FN}=M_H^{\rm K}+M_c^{\rm K}$, where $M_H^{\rm K}$ is the horizon Komar mass. For the PN metric, we identify $M_H^{\rm K}$ with the BH mass $M$ and determine $M_c^{\rm norm}$ by matching the cloud contribution to the asymptotic mass coefficient measured from the $g_{tt}$ component.

Subtracting the vacuum background with the same BH parameters
from Eq.~\eqref{eq:1pNmetric}, with $U^{\rm BH}=-M/r$, gives
\begin{equation}
\begin{aligned}
 \delta g_{tt}^{S}
 ={}&-2U^S-2\psi^S-2(U^S)^2+\frac{4M}{r}U^S,
\end{aligned}
\label{eq:gtt_cloud_PN}
\end{equation}
where
$\psi^S=\psi_E^S+\psi_{{\rm BH,int}}^S
+\psi_{{\rm S,int}}^S$. We define the corresponding PN cloud mass $M_c^{\rm PN}$ such that at large $r$:
\begin{equation}
 \delta g_{tt}^{S}
 =\frac{2M_c^{\rm PN}}{r}
 +\mathcal{O}(r^{-2}).
 \label{eq:gtt_cloud_asymptotic}
\end{equation}

For the $|211\rangle$ state,
$E_{n\ell}=-\mu\alpha^2/8$ and the potentials derived in
App.~\ref{app:pn_potentials} have the asymptotic forms
\begin{align}
 U^S
 &=-\frac{M_c^{\rm norm}}{r}
 +\mathcal{O}(r^{-3}),
 \label{eq:US_asymptotic}\\
 \psi_E^S
 &=\frac{\alpha^2M_c^{\rm norm}}{2r}
 +\mathcal{O}(r^{-3}),
 \label{eq:psiE_asymptotic}\\
 \psi_{{\rm BH,int}}^S
 &=-\frac{3\alpha^2M_c^{\rm norm}}{2r}
 +\mathcal{O}(r^{-3}),
 \label{eq:psiBH_matching_asymptotic}\\
 \psi_{{\rm S,int}}^S
 &=-\frac{711}{640}
 \frac{\alpha\mu(M_c^{\rm norm})^2}{r}
 +\mathcal{O}(r^{-3}).
 \label{eq:psiSelf_matching_asymptotic}
\end{align}
The terms $(U^S)^2$ and $MU^S/r$ decay as $r^{-2}$ and therefore do not affect the mass coefficient.

Substitution into Eq.~\eqref{eq:gtt_cloud_PN} therefore yields
\begin{equation}
\begin{aligned}
 M_c^{\rm PN}
 ={}&(1+\alpha^2)M_c^{\rm norm}+\frac{711}{640}\alpha\mu
 (M_c^{\rm norm})^2.
\end{aligned}
\label{eq:Mc_PN_physical}
\end{equation}
This relation refers to the prescribed hydrogenic profile and the
complete set of cloud potentials retained here.  In particular,
the normalization in Eq.~\eqref{eq:Mc_normalization} is not the
relativistic conserved particle number multiplied by $\mu$;
the sign of the correction in Eq.~\eqref{eq:Mc_PN_physical}
should not be interpreted as the sign of the binding energy
at fixed particle number.

When the PN background mass is identified with the FN horizon
Komar mass, $M=M_H^{\rm K}$, matching the total asymptotic masses
reduces to
\begin{equation}
 M_c^{\rm PN}=M_c^{\rm K}.
 \label{eq:Mc_matching_condition}
\end{equation}
Defining
\begin{equation}
 A=1+\alpha^2,
 \qquad B=\frac{711}{640}\alpha\mu,
\end{equation}
the nonnegative solution for the required normalization is
\begin{equation}
 M_c^{\rm norm}
 =\frac{2M_c^{\rm K}}
 {A+\sqrt{A^2+4BM_c^{\rm K}}}.
 \label{eq:Mc_normalization_matched}
\end{equation}

\bibliography{biblio}

\end{document}